\documentclass[12pt]{article}
\usepackage{epsfig,amssymb}
\usepackage{amsmath}
\usepackage{graphics,psfrag,rotating,cite}

\newcommand{\tr}{\mbox{tr}}

\def\pslash{{p\mkern-7mu/}}

\def\partials{{\partial \mkern-9mu/}}
\def\lsim{\raisebox{-.4ex}{$\stackrel{<}{\scriptstyle \sim}$\,}}
\def\gsim{\raisebox{-.4ex}{$\stackrel{>}{\scriptstyle \sim}$\,}}

\makeatletter
\renewcommand{\section}{\setcounter{equation}{0}\@startsection
    {section}%
    {1}%
    {0pt}%
    {-1\baselineskip}%
    {0.4\baselineskip}%
    {\large\bfseries}}%
\renewcommand{\subsection}{\@startsection
    {subsection}%
    {2}%
    {0pt}%
    {-0.75\baselineskip}%
    {0.2\baselineskip}%
    {\bfseries}}%
\renewcommand{\subsubsection}{\@startsection
    {subsubsection}%
    {3}%
    {0pt}%
    {-0.5\baselineskip}%
    {0.1\baselineskip}%
    {\sc}}%
\makeatother

\begin{document}
\input epsf \renewcommand{\topfraction}{0.8}
\pagestyle{empty} \vspace*{-5mm}
\hfill CERN-PH-TH/2006-097
\begin{center}
\Large{\bf On the electroweak symmetry breaking\\ in the Littlest Higgs
model} \\
\vspace*{2cm} \large{ Antonio Dobado, Lourdes Tabares} \\
\vspace{0.2cm} \normalsize
Departamento de  F\'{\i}sica Te\'orica I,\\
Universidad Complutense de
Madrid, E-28040 Madrid, Spain
\vspace{0.2cm}\\
and\vspace{0.2cm}\\
\large{Siannah Pe\~naranda}
\\ \vspace{0.2cm} \normalsize
CERN TH Division, Department of Physics,
CH-1211 Geneva 23, Switzerland\\
and IFIC - Instituto de F\'{\i}sica Corpuscular,
CSIC - Universitat de Val{\`e}ncia,\\ Apartado de Correos 22085,
E-46071 Valencia, Spain\\
\vspace*{2cm}{\bf ABSTRACT}
\end{center}
In the $SU(5)/SO(5)$ little Higgs models radiative corrections
give rise to the $SU(2)_L\times U(1)_Y$ symmetry breaking. In this
work we start a program for a detailed determination of the
relevant terms of the Higgs effective potential by computing the
contribution of the $b$, $t$ and $T$ quarks at the one-loop level,
as an starting point for higher-loops computation. In spite of the
fact that some two-loop level contributions are well known to be
important, we use our preliminary one-loop result to illustrate
that, by demanding the effective potential to reproduce exactly
the Standard Model Higgs potential, and in particular the relation
$m_H^2= 2 \lambda v^2= 2\mu^2$, it will be possible to set new
constraints on the parameter space of the Littlest Higgs model
when the computation of all the relevant contributions to the
Higgs effective potential is completed. \noindent

\newpage
\setcounter{page}{1} \pagestyle{plain} \textheight 20 true cm
\section{Introduction}

In the last years a lot of work has been devoted to the so-called
little Higgs  models (see~\cite{Schmaltz} and~\cite{review1} for
recent reviews). These models use an old suggestion by Georgi and
Pais~\cite{Georgi} in which the Higgs is assumed to be a (pseudo)
Goldstone boson associated to some global spontaneous symmetry
breaking~\cite{Georgi2}. For instance, in the case of the
paradigmatic Littlest Higgs (LH) \cite{Cohen}, a $SU(5)$ to
$SO(5)$ breaking is assumed to happen at some scale $f$. The Higgs
field is just one of the corresponding $14$ Goldstone bosons and
therefore it is, in principle, massless. The $SU(5)$ subgroup
$(SU(2)\times U(1))_{1} \times (SU(2)\times U(1))_{2}$ is gauged
so that the axial $(SU(2)\times U(1))_{1-2}$ is spontaneously
broken, the corresponding gauge bosons being typically heavy
($W'^{a}$ and $B'$). The diagonal $(SU(2)\times
U(1))_{1+2}=SU(2)_L\times U(1)_Y$ remains unbroken and corresponds
to the electroweak SM group. However, radiative corrections coming
from the fermionic sector of the model, mainly the third quark
generation and an additional vector-like quark $T,$ give rise to
an effective potential that produces a further spontaneous
symmetry breaking of the SM $SU(2)_L\times U(1)_Y$ down to
$U(1)_{\rm{em}}$. In this way some of the Goldstone bosons acquire
masses quadratic in the cut-off $\Lambda$, which is expected to be
of the order of $4\pi f$~\cite{relationlambdaf}, but the SM model
Higgs gets a mass that grows only as the logarithm of $\Lambda$ at
the one-loop level. This is the explanation in this setting of why
the Higgs is expected to be relatively light ($115$ GeV $< m_H <
200 $ GeV). The rest of the Goldstone bosons gives masses to the
different gauge bosons through the Higgs mechanism in the $SU(5)$
to $SO(5)$ or in the SM breaking. Thus, the LH model explains in a
natural and elegant way the expected low value of the Higgs boson
mass. In addition this model provides a very rich phenomenology,
which could be probed in the next-generation colliders such as the
LHC~\cite{Logan,Peskin}. Since the original proposal of the LH,
many other little Higgs versions have appeared~\cite{others}. Some
of these models try to improve the consistency and reduce the need
for fine-tuning in this kind of models (see~\cite{Casas} for
details). Specially interesting from the phenomenological point of
view is the LH version in which the $SU(5)$ gauged subgroup is
just $SU(2)_{1} \times SU(2)_{2} \times U(1)$. In this case, after
the first spontaneous broken symmetry, we have only three massive
gauge bosons associated to the $SU(2)_{1-2}$ group, $W'^{a}$, and
four massless gauge bosons, i.e. the SM gauge bosons
\cite{Peskin}.

Nevertheless, it is clear that any viable little Higgs model has
to fulfil  the basic requirement of reproducing the SM model at
low energies. This implies, in particular, not only to have the
proper low energy degrees of freedom, but also to reproduce the SM
model action as the low energy effective action of the LH model,
whenever one be near the physical minimum. In this work we compute
the contribution of the $t, b$ and $T$ quarks to the effective
potential for the SM Higgs doublet $H=(H^0,H^+)$, which gives rise
to the electroweak symmetry breaking in the LH model. The first
terms of this potential are found to have the standard form:
\begin{eqnarray}
\label{eq:potential}
V_{\rm{eff}}=-\mu^{2}HH^{\dag}+\lambda (HH^{\dag})^{2}\,,
\end{eqnarray}
with positive  $\mu^2$ and $\lambda$. Other relevant contributions
coming from gauge bosons, scalars and other higher loops are
expected to go in the opposite direction, but they are also
supposed to have a smaller absolute value so that they do not
change the $\mu^2$ and $\lambda$ signs. The $\mu^2$ sign and value
are well known \cite{Cohen,Peskin}, and effectively they are the
right ones to produce the electroweak symmetry breaking, giving a
Higgs mass $m_H^2=2 \mu^2$. However, the full expression for
$\lambda$ has not
  been analyzed in detail. Several relations for the threshold
  corrections to this parameter in the presence of a $10$ TeV cut-off,
  depending of the UV-completion of the theory, has been reported before
  (see, for example~\cite{italianos}). The radiative corrections to
  $\lambda$, at the one-loop level, have not been computed so far.

The computation of the $\lambda$ parameter is important for
several reasons. First, it must be positive, for the low energy
effective action to make sense (otherwise the theory would not
have any vacuum). In addition, from the effective potential above,
one gets the simple formula $m^2_H=2 \lambda v^2$ or,
equivalently, $\mu^2 = \lambda v^2$, where $v$ is the SM vacuum
expectation value ($H=(0,v)/\sqrt{2}$), which is set by experiment
(for instance from the muon lifetime) to be $v\simeq 245$ GeV. By
computing the effective action by the Higgs doublet in the context
of the LH model, taking into account the $t,b$ and $T$ quarks
only, i.e. the modes responsible for the electroweak symmetry
breaking, it is possible to obtain $\mu^2$ and $\lambda$ in terms
of the $\lambda_T, f$ and $\Lambda$ parameters of the LH model,
where $\lambda_T$ is the $T$ Yukawa coupling, $f$ is the scale of
the $SU(5)/SO(5)$ symmetry breaking, and $\Lambda$ is the
ultraviolet cut-off (in fact $\lambda$ has also a small dependence
on an infrared cut-off $m\sim v$). In other words, we can find the
functions:
\begin{eqnarray}
\mu^2=\mu^2\,(\lambda_T, f, \Lambda) \,, \nonumber  \\
\lambda=\lambda\,(\lambda_T, f, \Lambda)\,.
\end{eqnarray}
As is well known, $\mu$ depends on the logarithm of $\Lambda$ at one-loop level,
but $\lambda$ has also a much stronger quadratic dependence on
this cut-off. Moreover, according to the previous discussion, the
consistency of the low energy theory sets the following highly
non-trivial constraint on the LH model parameters:
\begin{equation}
\mu^2(\lambda_T,f,\Lambda) = \lambda(\lambda_T,f,\Lambda)v^2+...
\,, \label{eq:relation}
\end{equation}
where the periods include corrections coming from gauge bosons,
scalars and other higher order loops.

In this work we compute the contribution to these functions coming
from  the $t, b$ and $T$ quarks present in the LH model at the
one-loop level which are the relevant ones for having symmetry
breaking. Then we use the result to illustrate the kind of bounds
and restrictions that must be set on the LH model fermion
parameters in order to obtain the SM potential from the effective
Higgs potential which should also include gauge, scalar and other
higher loop contributions \cite{future}. This analysis is crucial
if one assumes that the new physics decouples from the low energy
scale.

The outline of the paper is as follows: In Section 2 we review
briefly the LH model and set the notation we are going
to use. In Section 3 we compute the effective action for a
constant SM Higgs doublet, i.e. the effective potential, by using
standard techniques (see for example~\cite{book}), and we obtain the
$\mu=\mu(\lambda_T,f,\Lambda)$ and
$\lambda=\lambda(\lambda_T,f,\Lambda)$ functions. Section 4 is
devoted to the study of the above-mentioned constraints that our
computation sets on the LH model parameter space and,
finally, in Section 5 we summarize our main results and present some
conclusions and remarks.

\section{The model}

The LH model is based on the assumption that there is a physical
system with a global $G=SU(5)$ symmetry, which is spontaneously
broken to a $H=SO(5)$ symmetry at a high scale $\Lambda$ through a
vacuum expectation value of order $f$. Thus the spectrum of the
theory will contain in principle $14$ Goldstone bosons including
the SM complex doublet $H=(H^{0},H^{+})$. In addition, the $SU(5)$
subgroup $(SU(2)\times U(1))_{1} \times (SU(2)\times U(1))_{2}$ is
gauged, its diagonal subgroup $(SU(2)\times U(1))_{1+2}$ being the
SM electroweak group $SU(2)_L \times U(1)_Y$. This group remains
unbroken after the $SU(5)$ breaking to $SO(5)$ and consequently
the electroweak gauge bosons $W_{\mu}^a$ and $B_{\mu}$ are
massless at this level. However, the $(SU(2)\times U(1))_{1-2}$
group becomes spontaneously broken
 and the corresponding gauge bosons  $W_{\mu}^{'a}$ and $B'_{\mu}$ get masses
 of order $f$  through the Higgs mechanism.
Each of these two gauge groups must commute with a different
subgroup $SU(3)$ that acts non-linearly on the Higgs, i.e. when
both weak gauge interactions are included, the Higgs is a
pseudo-Goldstone boson whose mass is protected by the underlying symmetry
but, conversely, if just one of these interactions is considered,
the $SU(3)$ symmetry is recovered~\cite{Cohen}.

With the global $SU(5)$ symmetry breaking into its subgroup
$SO(5)$, we have $14$ Goldstone bosons, which transform under the
electroweak group as a real singlet, a real triplet, a complex
doublet and a complex triplet.  The real singlet and the real
triplet become the longitudinal part of the $B'_{\mu}$ and
$W_{\mu}^{'a}$ bosons through the Higgs mechanism, and the last
two Goldstone boson multiplets can be interpreted as the SM Higgs
doublet and an additional complex triplet, i.e. we still have $10$
massless Goldstone bosons. These particles will get radiative
masses after the introduction of appropriate gauge and Yukawa
couplings to the third-generation $b$ and $t$ quarks and an
additional vector-like quark $T$, the Yukawa contributions being
responsible to give the expected sign to the Higgs doublet mass.
Then, the magic of the model produces a Higgs mass, which is
quadratically divergence-free at the one-loop level. In this way
we obtain a light Higgs in a natural way, thanks to the pseudo-Goldstone
boson nature of this field. The complex triplet is not protected
in the same way and quark radiative corrections make it typically
much more massive, thus evading the experimental constraints.

According to the previous discussion, the low energy dynamics of
the LH model can be described by a $(SU(2)\times U(1))_{1}
\times (SU(2)\times U(1))_{2}$ gauged non-linear
 sigma model based on the coset $K=G/H=SU(5)/SO(5)$
(see for instance~\cite{book}). The Goldstone
 boson fields can be arranged in a  $5 \times 5$ matrix $\Sigma$ given by:
 \begin{equation}
\Sigma=e^{2 i\Pi/f} \Sigma_{0},
\end{equation}
where:
\begin{equation}
\Sigma_{0}= \left(%
\begin{array}{ccc}
  0 & 0 & \textbf{1} \\
  0 & 1 & 0 \\
  \textbf{1} & 0 & 0 \\
\end{array}%
\right)
\end{equation}
has the proper $SU(5)$ symmetry breaking structure,
 $\textbf{1}$ being the $2 \times 2$ unit matrix, and
\begin{equation}
\Pi=\left(%
\begin{array}{ccc}
\xi & \frac{-i}{\sqrt{2}}H^{\dag} & \phi^{\dag} \\
  \frac{i}{\sqrt{2}}H & 0 & \frac{-i}{\sqrt{2}}H^{*} \\
  \phi & \frac{i}{\sqrt{2}}H^{T} & \xi^{T} \\
\end{array}%
\right)+\frac{1}{\sqrt{20}}\,\eta\, {\mbox{diag }}(1,1,-4,1,1)
\end{equation}
is the Goldstone bosons matrix, with $H=(H^{0},H^{+})$ the
SM Higgs doublet, $\eta$ being the real scalar, and
$\xi$ and $\phi$ encoding the real triplet and the complex triplet
respectively:
\begin{equation}
\xi=\left(%
\begin{array}{cc}
 \frac{1}{2} \xi^{0} & \frac{1}{\sqrt{2}}\xi^{+} \\
  \frac{1}{\sqrt{2}}\xi^{-} & -\frac{1}{2}\xi^{0}
\end{array}%
\right)
~~~~~~~~~~~{\mbox{ and }}~~~~~~~~~~~~
\phi=\left(%
\begin{array}{cc}
  \phi^{0} & \frac{1}{\sqrt{2}}\phi^{+} \\
\frac{1}{\sqrt{2}}\phi^{+} & \frac{1}{\sqrt{2}}\phi^{++}
\end{array}%
\right)\,.
\end{equation}

The Lagrangian of
the gauged non-linear sigma model is given by:
\begin{equation}
{\cal L}_0 = \frac{f^2}{8}\tr D_{\mu}\Sigma (D^{\mu}\Sigma)^\dag \,,
\end{equation}
where the covariant derivative is defined as~\cite{Cohen}:
\begin{equation}
D_{\mu}\Sigma=\partial_{\mu}\Sigma-i\sum_{j=1}^{2}g_jW^a_j(Q_j^a\Sigma +\Sigma Q_j^{aT})
-i\sum_{j=1}^{2}g'_jB_j(Y_j\Sigma+\Sigma Y_j^{T})\,,
\end{equation}
where $g_j$ and $g'_j$ are the gauge couplings,
$Q_{1ij}^a=\sigma_{ij}^a/2$ for $i,j=1,2$,
$Q_{2ij}^{a}=\sigma_{ij}^{a*}/2$ for
 $i,j=4,5$, and zero otherwise, $Y_1=$ diag$(-3,-3,2,2,2)/10$ and
$Y_2=$ diag$(-2,-2,-2,3,3)/10$.
By diagonalizing the gauge boson mass matrix contained in this
Lagrangian one gets the massless $W$ and $B$ SM bosons and the
massive $W'$ and $B'$ gauge bosons mentioned above as:
\begin{eqnarray}
W^a   & = & c_{\psi} W_1^a+ s_{\psi} W_2^a \nonumber\\
 W^{'a}  & = & s_{\psi} W_1^a- c_{\psi} W_2^a \,,
\end{eqnarray}
where
\begin{eqnarray}
s_{\psi} &=& \sin \psi = \frac{g_1}{\sqrt{g_1^2+g_2^2}} \nonumber\\
 c_{\psi} &=& \cos \psi =
\frac{g_2}{\sqrt{g_1^2+g_2^2}}
\end{eqnarray}
with $M_W= 0$  and  $M_{W'}= f \sqrt{g_1^2+g_2^2}/2$. In a similar
way we have
\begin{eqnarray}
B  & = & c'_{\psi} B_1+ s'_{\psi} B_2 \\ \nonumber
 B'  & = & s'_{\psi} B_1- c'_{\psi} B_2 \,,
\end{eqnarray}
where
\begin{eqnarray}
s'_{\psi}&=&  \sin \psi' = \frac{g'_{1}}{\sqrt{{g'}_{1}^{\,2}+{g'}_{2}^{\,2}}}\\
\nonumber
 c'_{\psi}&=& \cos \psi' =
\frac{{g'}_{2}}{\sqrt{{g'}_{1}^{\,2}+{g'}_{2}^{\,2}}}\,,
\end{eqnarray}
where $M_B=0$ and $M_{B'}= f \sqrt{g_1^{'2}+g_2^{'2}}/\sqrt{20}$.

A modified version of the LH models, such that the gauge
subgroup of $SU(5)$ is $[SU(2) \times SU(2)\times U(1)_Y]$ rather than
$[SU(2)\times U(1)_Y]^2$, has also been introduced~\cite{Peskin}.
In this case, the covariant derivative is defined as:
\begin{equation}
D_{\mu}\Sigma=\partial_{\mu}\Sigma
-i\sum_{j=1}^{2}g_jW^a_j(Q_j^a\Sigma +\Sigma Q_j^{aT})
-i g' B (Y \Sigma+\Sigma Y^{T})\,,
\end{equation}
where the generators $Q_{j}^{a}$ are the same as in the previous case,
and $Y=\frac{1}{2}$diag$(-1,-1,0,1,1)$. The field content of the matrix
$\Pi$ in $\Sigma$ is the same as in the LH model but there is no $B^{'}$
now. This model will be considered in Section 4 when some phenomenological
consequences of considering the gauge sector are discussed.

Then, at the tree level, the $SU(2)_L\times U(1)_Y$ SM gauge group
remains unbroken. The spontaneous symmetry breaking of this group
is produced in this model radiatively mainly by the quark loops
 from the third generation, which will be initially denoted by $u$
and $t$ and the additional vector-like quark denoted by $U.$ The
interactions between these fermions and  the Goldstone bosons are
given by the Yukawa Lagrangian:
\begin{equation}
\textit{L}_{\rm{Yuk}}=-\frac{\lambda_{1}}{2}f\,
\overline{u}_{R}\,\epsilon_{mn}\,\epsilon_{ijk}\,\Sigma_{im}\,
\Sigma_{jn}\,\chi_{Lk}-\lambda_{2}\,f\, {\overline{U}}_{R}\,U_{L}+\mbox{h.c.},
\label{lagran}
\end{equation}
where $m,n=4,5$, $i,j=1,2,3$, and
\begin{eqnarray}
\overline{u}_{R}&=& c \,\overline{t}_{R}+ s\,  {\overline{T}}_{R}\,,\nonumber\\
{\overline{U}}_{R}&=&-s \,\overline{t}_{R}+   c \,{\overline{T}}_{R},
\end{eqnarray}
with:
\begin{eqnarray}
c&=&\cos
\theta=\frac{\lambda_{2}}{\sqrt{\lambda_{1}^{2}+\lambda_{2}^{2}}},\nonumber\\
s&=&\sin \theta =
\frac{\lambda_{1}}{\sqrt{\lambda_{1}^{2}+\lambda_{2}^{2}}}\,,
\end{eqnarray}
and
\begin{equation}
\chi_{L}=\left(%
\begin{array}{c}
  u \\
  b \\
  U \\
\end{array}%
\right)_{L}=\left(%
\begin{array}{c}
  t \\
  b \\
  T \\
\end{array}%
\right)_{L}.
\end{equation}
Here $t\,, b$, $T$ are the mass eigenvectors coming from the mass
matrix included in the Yukawa Lagrangian with eigenvalues:
$m_{t}=m_b=0$ and $m_{T}=f\sqrt{\lambda_{1}^{2}+\lambda_{2}^{2}}$.
Thus the quark $t$ is massless and it acquires mass only when the
electroweak symmetry is broken, contrary to the quark $T$, which is
massive already at this level.

Then, the Yukawa Lagrangian can be written as:
\begin{equation}
\textit{L}_{\rm{Yuk}}=\overline{\chi}_{R}\, \hat{I}_{3x3}\,
\chi_{L}+\mbox{h.c.}\,,
\end{equation}
$\hat{I}_{3x3}$ being the interaction matrix defined below, and
$$\chi_{R}=\left(%
\begin{array}{c}
  t \\
  b \\
  T \\
\end{array}%
\right)_{R}\,.$$ Since we are interested in the computation of the
fermion contribution to the SM Higgs effective potential, we
set $\xi= \phi = \eta = 0 $ and thus, the interaction matrix
$\hat{I}$ is given by:
\begin{eqnarray}
\hat{I}=\left(%
\begin{array}{ccc}
 -\sqrt{2}\lambda_{1}cH^{0}\Theta
 &-\sqrt{2}\lambda_{1}cH^{+}
 \Theta &\lambda_{1}c \frac{HH^{\dag}}{f}\Theta^{'}\\
0 & 0 & 0 \\
-\sqrt{2}\lambda_{1}sH^{0}\Theta & -\sqrt{2}\lambda_{1}sH^{+}
\Theta &\lambda_{1}s \frac{HH^{\dag}}{f} \Theta^{'}
\end{array}%
\right),
\end{eqnarray}
where $\Theta$ and $\Theta^{'}$ are functions on $H H^\dag/ f^2$
whose expansion starts as:
\begin{eqnarray}
\Theta\left(\frac{HH^{\dag}}{f^2}\right)=1-\frac{2 HH^{\dag}}{3 f^{2}}+...\\
\nonumber
\Theta'\left(\frac{HH^{\dag}}{f^2}\right)=1-\frac{HH^{\dag}}{3f^{2}}+...
\end{eqnarray}
Therefore the complete Lagrangian for the quarks is:
\begin{eqnarray}
\textit{L}_{\chi}&=&\textit{L}_{0}+\textit{L}_{\rm{Yuk}}=
\overline{\chi}_{R}(i
\partials-M+\hat{I})
\chi_{L}+\mbox{h.c.}
\end{eqnarray}
with $M=$diag$(0,0,m_T)$.

In the LH model the electroweak symmetry breaking is produced
mainly by the three quarks included in the above Lagrangian, whilst
the gauge bosons and the complex triplet tend to restore the
symmetry. In the following we will consider only the effect of the
quarks on the Higgs effective potential by turning off $g_1$ and $g_2$.

\section{The Higgs effective action and potential}

In order to compute the leading fermion contribution to the Higgs
effective potential we will start from the Higgs effective action
obtained from the $t$, $b$ and $T$ quarks at the one-loop level,
which is an exact computation in this case since the action is
quadratic on these fields. Thus this effective action is given by:
\begin{equation}
e^{iS_{\rm{eff}}[H]}=\int [d\chi] [d\overline{\chi}] e^{i
S[H,\chi,\overline{\chi}]}\,,
\end{equation}
with
\begin{equation}
S[H,\chi,\overline{\chi}]=\int d^{4}x \,(\partial_{\mu}H
\partial^{\mu}H^{\dag}+\textit{L}_{\chi}).
\end{equation}
By using  standard techniques (see for instance \cite{book}) we obtain
 the following result for the effective action,
\begin{equation}
S_{\rm{eff}}[H]=\int d^{4}x \partial_{\mu}H \partial^{\mu}H^{\dag}+
\Delta S,
\end{equation}
with
\begin{equation}
\Delta S=-i \mbox{Tr}\log(i \partials -M+\hat{I})=-i
\mbox{Tr}(1+G\hat{I})\,,
\end{equation}
where we have neglected a constant, irrelevant for the computation
of the effective action. The operator $G=(i\partials -M)^{-1}$ is
just the propagator for the free quarks, which is given by
\begin{equation}
G^{ab}(x,y)\equiv \int d\widetilde{k}\,e^{-ip(x-y)}
(\pslash-M)^{-1}_{ab}.
\end{equation}
Here $d\widetilde{k}\equiv d^4 k/(2\pi)^4$. By expanding the
logarithm, the effective action can be written as
\begin{equation}
\Delta S=-i \mbox{Tr}\Sigma_{k=1}^{\infty}\frac{(-1)^{k+1}}{k}
(G\hat{I})^{k}=\Sigma_{k=1}^{\infty} \Delta S^{(k)}.
\end{equation}
Now in order to obtain the effective potential we have only to
consider constant Higgs fields, i.e. we set $\partial_{\mu}H=0$.
Thus we have:
\begin{equation}
\Delta S|_{H={\rm{const.}}}=-\int d^{4}x\, V_{\rm{eff}}(H)\,.
\label{eq:Hpotential}
\end{equation}
In the following we will take $H$ as a constant. The effective
potential can be computed as a power series of $HH^{\dag}/f^2$,
with arbitrary higher powers of this parameter. However, in order
to produce the electroweak symmetry breaking, it is sufficient to
compute just the first two terms of this expansion. Thus the
effective potential can be written as given in~(\ref{eq:potential}).

\begin{figure}[t]\begin{center}
\epsfig{file=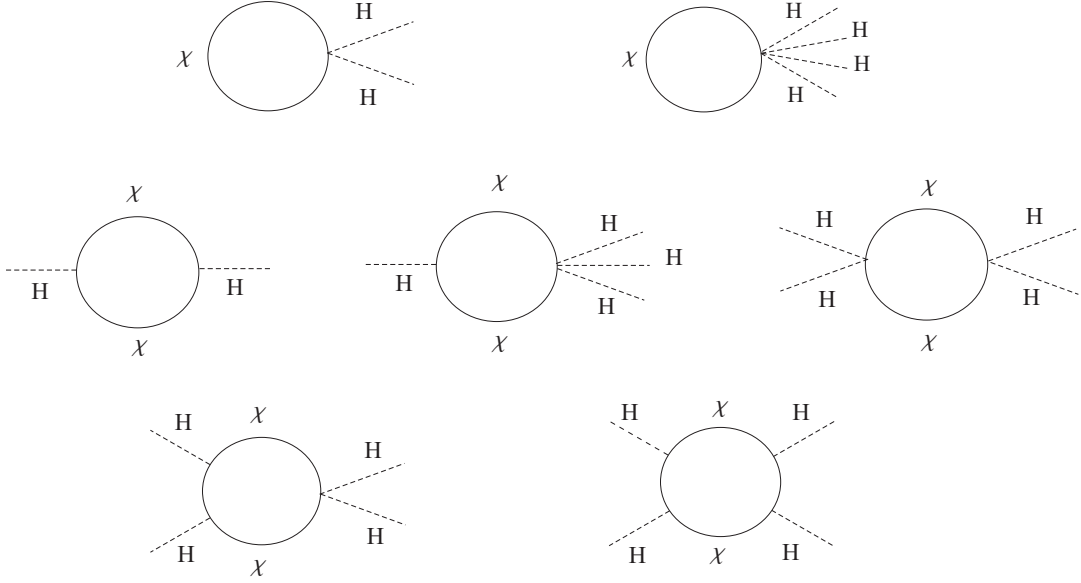,scale=0.77}
\caption{One-loop diagrams. $H=(H^{0},H^{+})$ and $\chi \equiv t, b, T$,
  with all possible combinations of these particles in the loop diagrams.}
\label{diagrams}\end{center}
\end{figure}

It is then not difficult to see that the computation of the
$\mu^2$ and $\lambda$ parameters requires $\Delta S$ to be considered
up to the fourth term. The generic one-loop diagrams that must be computed are
shown in Fig.~\ref{diagrams}.
By using well known methods it is straightforward to obtain the
different contributions after some work. The first one ($k=1$)
corresponds to the first two diagrams in Fig.~\ref{diagrams}
and it is given by:
\begin{equation}
\Delta S^{(1)}[H]=-i
\mbox{Tr}(G\hat{I})=-\frac{4}{3}\lambda_{1}\,m_{T}\,f\, s \int
d^{4}x\left(\frac{HH^{\dag}}{f^{2}}-\frac{(HH^{\dag})^{2}}{f^{4}}\right)
I_{0}(m_{T}^{2})\,,
\end{equation}
where the divergent integral $I_{0}(m_{T}^{2})$ is:
\begin{equation}
I_{0}(m_T^{2})=\int d \tilde{p}\frac{i}{p^{2}-m_{T}^{2}}\,.
\end{equation}
In order to make sense of this integral we will use an ultraviolet
cut-off $\Lambda$, where our effective description of
the low energy dynamics breaks down. The result is:
\begin{equation}
 I_{0}(m_T^{2})=\frac{1}{(4\pi)^{2}}\left[\Lambda^{2}-m_T^{2}
\log\left(\frac{\Lambda^{2}}{m_T^{2}}+1\right)\right]\,.
\end{equation}
For $k=2$ (see the three generic diagrams on the second line of
Fig.~\ref{diagrams}), one gets:
\begin{eqnarray}
\Delta S^{(2)}[H]&=&\frac{i}{2} \mbox{Tr}(G\hat{I})^{2}=4
\lambda_{1}^{2}\int d^{4}x HH^{\dag}(c^{2} I_{0}(0)+s^{2}
I_{0}(m_{T}^{2}))\nonumber \\
&+&2\lambda_{1}^{2}\int
d^{4}x\frac{(HH^{\dag})^2}{f^{4}}(I_{0}(0)+ 2
m_{T}^{2}s^{2} I_{1}(m_{T}^{2}))   \nonumber \\
& - & \frac{16}{3}\lambda_{1}^{2}\int
d^{4}x\frac{(HH^{\dag})^2}{f^{4}}(c^{2} I_{0}(0)+s^{2}
I_{0}(m_{T}^{2}))\,,
\end{eqnarray}
where the new divergent integral $I_{1}(m_T^{2})$ properly
regularized is given by:
\begin{equation}
I_{1}(m_T^{2})=\int d
\tilde{p}\frac{i}{(p^{2}-m_{T}^{2})^{2}}=-\frac{1}{(4\pi)^{2}}
\left[\log\left(\frac{\Lambda^{2}}{m_T^{2}}+1\right)-
\frac{1}{1+\frac{m_T^{2}}{\Lambda^{2}}}\right]\,.
\end{equation}
The $k=3$ contribution (see the first diagram in the bottom line of
Fig.~\ref{diagrams}) is:
\begin{equation}
\Delta S^{(3)}[H]=-\frac{i}{3} \mbox{Tr}(G\hat{I})^{3}=-8m_{T}
\lambda_{1}^{3}s\int d^{4}x \frac{(HH^{\dag})^2}{f}(c^{2}
I_{2}(m_{T}^{2})+s^{2} I_{1}(m_{T}^{2}))\,,
\end{equation}
where the divergent integral $I_{2}(m_T^{2})$ can be written as:
\begin{equation}
I_{2}(m_T^{2})=\int d
\tilde{p}\frac{i}{p^{2}(p^{2}-m_{T}^{2})}=
-\frac{1}{(4\pi)^{2}}\log\left(\frac{\Lambda^{2}}{m_T^{2}}+1\right)\,.
\end{equation}
Finally, for $k=4$, we get
\begin{equation}
\Delta S^{(4)}[H]=\frac{i}{4} \mbox{Tr}(G\hat{I})^{4}=
4\lambda_{1}^{4}\int d^{4}x
(HH^{\dag})^{2}(s^{4}I_{1}(m_{T}^{2})+2s^{2}c^{2}I_{2}(m_{T}^{2})+c^{4}I_{2}(0)).
\end{equation}
Here we need to compute $I_{2}(0).$ This integral is not only
ultraviolet-divergent but also infrared-divergent. Thus we need to
introduce a new infrared cut-off $m$ (obviously the natural value
for this cut-off is of the order of $v,$ i.e. the scale of the
electroweak symmetry breaking). Then we find:
\begin{equation}
I_{2}(0)=\int d
\tilde{p}\frac{i}{p^{4}}=-\frac{1}{(4\pi)^{2}}
\log\left(\frac{\Lambda^{2}}{m^{2}}\right)\,.
\end{equation}

Therefore, by using the previous results, it is possible to write
the Higgs effective potential parameters as~{\footnote{Our results agree with
    previous ones for $\mu^2$ (see, for
example,~\cite{review1} and references therein).}}:
\begin{equation}
\mu^{2}= N_{c} \frac{m_{T}^{2} \lambda_{t}^{2}}{4 \pi^{2}}
\log\left(1+\frac{\Lambda^{2}}{m_{T}^{2}}\right),
\label{eq:muparameter}
\end{equation}
and
\begin{eqnarray}
\lambda&=&\frac{N_{c}}{(4
\pi)^{2}}\left[2(\lambda_{t}^{2}+\lambda_{T}^{2})
\frac{\Lambda^{2}}{f^{2}}-\log\left(1+\frac{\Lambda^{2}}{m_{T}^{2}}\right)
\left(-\frac{2m_{T}^{2}}{f^{2}}\left(\frac{5}{3}
\lambda_{t}^{2}+\lambda_T^{2}\right)+4\lambda_{t}^{4}+4(\lambda_{T}^{2}+
\lambda_{t}^{2})^{2}\right)\right.\nonumber\\
&&\left.-4\lambda_{T}^{2}\frac{1}{1+\frac{m_{T}^{2}}{\Lambda^{2}}}\left
(\frac{m_{T}^{2}}{f^{2}}-2\lambda_{t}^{2} -\lambda_{T}^{2}\right)
-4\lambda_{t}^{4}\log\left(\frac{\Lambda^{2}}{m^{2}}\right)\right],
\label{eq:lambdaTOP}
\end{eqnarray}
where $N_c$ is the number of colors and $\lambda_t$ and
$\lambda_T$ are, respectively, the SM top
Yukawa coupling and the heavy top Yukawa coupling,
given by~{\footnote{Here we assume that $m_t = \lambda_{t} v$.}}:
\begin{eqnarray}
\lambda_{t}&=&
\frac{\lambda_{1}\lambda_{2}}{\sqrt{\lambda_{1}^{2}+\lambda_{2}^{2}}}
\nonumber\\
\lambda_{T}&=&\frac{\lambda_{1}^{2}}{\sqrt{\lambda_{1}^{2}+\lambda_{2}^{2}}}\,.
\end{eqnarray}

There are several comments about this result which it is worthwhile
stressing. First, the effective potential depends on $H$ through the
combination  $|H|^2 = H H^\dagger $, thus reflecting the
fact that the radiative corrections considered preserve the SM
 $SU(2)_L\times U(1)_Y$ symmetry. However, $\mu^2 > 0$ and $\lambda >0$,
which are the right signs for these corrections to spontaneously break
this symmetry down to $U(1)_{\rm{em}}$.
 Thus the minima of the effective potential
 occur whenever $|H|^2= v^2/2\equiv \mu^2/(2\lambda)$. By
 choosing  the new vacuum as the state $H=(0,v)/\sqrt{2}$, we
 recover the above-mentioned SM symmetry breaking. In particular
 we find that the physical Higgs boson mass is given by $m_H^2= 2
 \lambda v^2$. Notice that for $\lambda <0$ the model would
be inconsistent and for $\mu^2 < 0$
 there would be no spontaneous symmetry breaking.

In spite of having found apparently the same results as in the
SM, there are however a number of nice properties in the LH model.
First of all, as we have shown,  the
Higgs potential parameters can be computed in terms of other more
fundamental parameters; therefore do not need to be introduced ad
hoc in order to get the appropriate symmetry breaking, as
happens in the SM. Moreover, the symmetry breaking appears as a
result of the dynamics, through third generation quarks radiative
corrections and not as a tree-level consequence of the SM
Lagrangian. On the other hand, the Higgs mass can be written as
$m_H^2=2\mu^2$ and $\mu^2$ is only a logarithmic-divergent
quantity. Therefore, the Higgs mass is, not only light as the
precision test of the SM seems to suggest, but also free from
the undesired quadratic divergences that appear in the original
formulation of the SM. Notice also that the quadratic divergences
appearing in $\lambda$ do not alter this result.

Once the spontaneous breaking of the SM symmetry is produced, the
Yukawa Lagrangian given above gives rise to a new mass matrix for
the $t$ and $T$ quarks (the $b$ quark remains massless). This mass
matrix can be diagonalized through a rotation of the left chiral
states $t_L$ and $T_L$ given by the angle $\theta_L$, which, for
$v\ll f,$ can be written as:
\begin{equation}
\sin\theta_{L} \simeq
\left(\frac{\lambda_{1}^{2}}{\lambda_{1}^{2}+\lambda_{2}^{2}}\right)
\frac{v}{f}
\end{equation}
and, by other rotation of the right states $t_R$ and $T_R$, given by
the angle $\theta_R$
\begin{equation}
\sin \theta_R \simeq s
\left[1-c^{2}\left(\frac{1}{2}-\left(
\frac{\lambda_{1}^{2}}{\lambda_{1}^{2}+\lambda_{2}^{2}}\right)\right)
\frac{v^{2}}{f^{2}}\right].
\end{equation}
After these rotations have been done the new mass eigenvalues
become:
\begin{equation}
m_t= \lambda_t \,v \left[ 1+ \left(-\frac{1}{3}+
\frac{1}{2}\frac{\lambda_{1}^{2}}{\lambda_{1}^{2}+\lambda_{2}^{2}}
\left(1-\frac{\lambda_{1}^{2}}{\lambda_{1}^{2}+\lambda_{2}^{2}}\right)\right)
\frac{v^{2}}{f^{2}}\right]
\end{equation}
and
\begin{equation}
 m_T=\frac{\lambda_t^2+\lambda_T^2}{\lambda_T}f
-\frac{1}{2}\frac{\lambda_{1}^{2}}{\sqrt{\lambda_{1}^{2}+\lambda_{2}^{2}}}
\left(1-\frac{\lambda_{1}^{2}}{\lambda_{1}^{2}+\lambda_{2}^{2}}\right)\frac{v^{2}}{f}.
\label{eq:afterbreak}
\end{equation}

\section{Constraints on the LH parameter space}

Whatever model one considers as a candidate for physics beyond
the SM, the consistency with present experimental data is a key
prediction of that candidate model. It is well known that indirect
constraints from precision electroweak measurements on new physics
at the TeV scale are severe. There exist several studies of the
corrections to electroweak precision observables in the Little
Higgs models, exploring whether there are regions of the parameter
space in which the model is consistent with
data~\cite{Schmaltz,review1,Logan,LoganP,EWPO1,EWPO2,recentpheno}.

 The effective Higgs potential
parameters are related with the vacuum expectation value $v$
through $\mu^2=\lambda v^2$; with $v \simeq 245$ GeV. By imposing
this condition, we could extract crucial information on the
allowed region of the parameter space in the LH model. To show
this we will consider here the case of the heavy quark
contribution to the Higgs potential. In spite of the fact that
other contributions coming from gauge bosons, scalars and other
higher loops are relevant, the fermionic sector provides a good
illustration of the kind of constraints that it will be possible
to set on the LH parameter space from the complete effective Higgs
potential.

\psfrag{mT}{{{\rotatebox{90}{$m_T$ [TeV]}}}}
\psfrag{mTf1}{{$m_T$ [TeV]}}
\psfrag{f}{{{\rotatebox{60}{$f$ [TeV]}}}}
\psfrag{lt}{{$\lambda_T$}}
\begin{figure}[t]
\begin{center}
\begin{tabular}{cc}
\epsfig{file=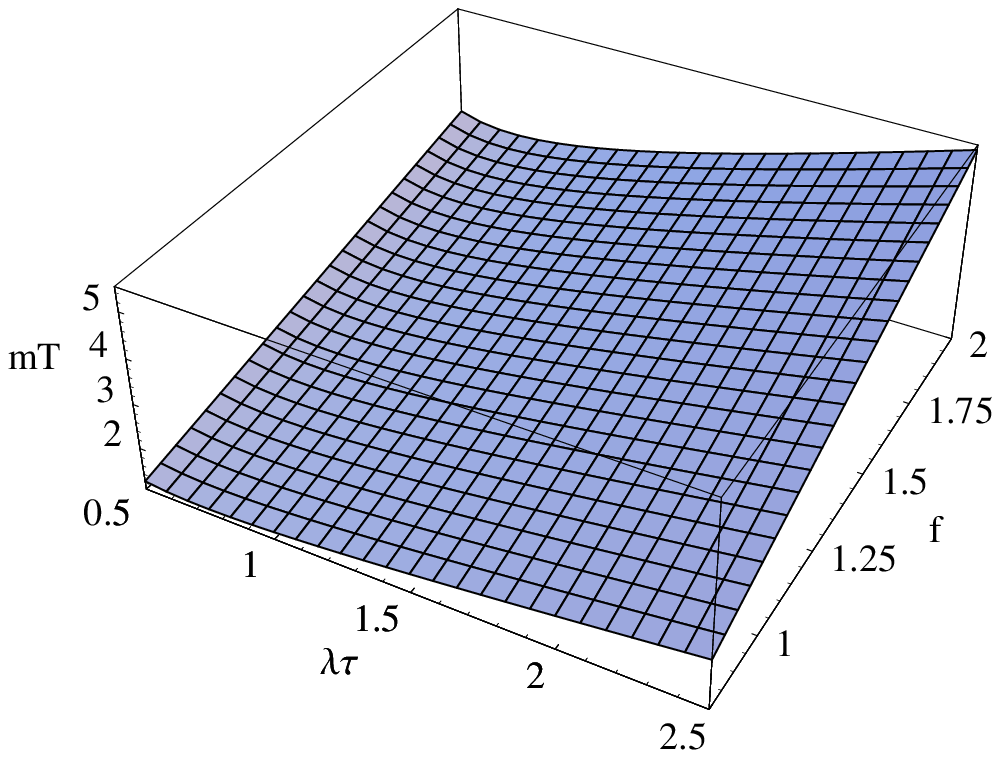,scale=0.8}&
\epsfig{file=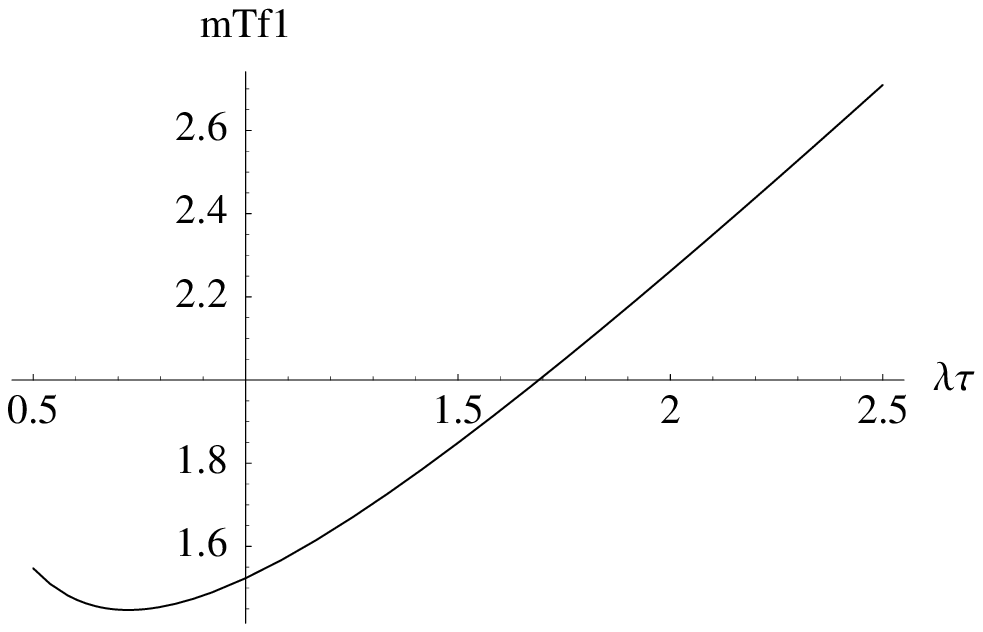,scale=0.8}\\
{\bf{(a)}} & {\bf{(b)}}
\end{tabular}
\end{center}\vspace*{-0.3cm}
\caption{{\bf{(a)}} $m_{T}$ as a function of $\lambda_T$ and $f$,
with $0.5 < \lambda_T < 2.5$ and $0.5<f<2$ and {\bf{(b)}} $m_{T}$ as a
function of $\lambda_T$ for $f=1$ TeV.}
\label{figuno}
\end{figure}

The contributions from the fermion sector of the model to $\mu^2$
and $\lambda$ are summarized in~(\ref{eq:muparameter}) and
(\ref{eq:lambdaTOP}), respectively; the undetermined parameters
of the model are the heavy top mass $m_{T}$, the coupling constant
$\lambda_T$, the symmetry breaking scale $f$, and the scale
$\Lambda$. However, there
are several relations between them which are worth remarking on.
Firstly, before the electroweak symmetry breaking by radiative
corrections, we have:
\begin{equation}
m_{T}=f \,\frac{\lambda_t^2+\lambda_T^2}{\lambda_T}\,,
\label{eq:canceldiv}
\end{equation}
which could, in principle, be tested by the LHC
experiments~\cite{review1,Peskin}. This relation is crucial for
the cancellation of quadratic divergences contributions to the
Higgs boson mass. Besides, considering $m_{T}$ much larger than
$2$ TeV would imply a large amount of fine-tuning in the Higgs
potential, and thus the heavy top should be below about $2$ TeV
($m_T\lsim 2.5$ TeV)~\cite{Cohen,Peskin}. Note that, since the
top-quark mass is already known in the SM, absolute bounds are
derived on the couplings, $\lambda_1, \lambda_2 \geq m_t/v$ or
$\lambda_1 \lambda_2 \geq 2 (m_t/v)^2$ ~\cite{Logan}.  As a
consequence, we get the bound $\lambda_T \gsim 0.5$, which have
been considered in our analysis. For the purpose of illustration,
the dependence of $m_{T}$ with $\lambda_T$ and $f$ is shown in
Fig.~\ref{figuno}{{a}} for $0.5 < \lambda_T < 2.5$ and $0.8$ TeV
$<f<2$ TeV. The corrections decrease with $\lambda_T$ having a
minimum for a value of the coupling constant closed to $1$, and
then they increase. Clearly, $m_{T}$ grows linearly with the
parameter $f$, $f\lsim 1$ TeV being the favored values at this
level, because of the condition $m_T\lsim 2$ TeV. For $\lambda_T >
1.7$ we get values of the heavy top mass above $2$ TeV, when $f=1$
TeV (Fig.~\ref{figuno}{{b}}). We note that once the spontaneous
symmetry breaking of the SM is produced, the heavy top mass is
reduced by ${\cal{O}}(v^2/f^2)$ terms (see
eq.~(\ref{eq:afterbreak})), but the reduction is only of about
$0.01$ TeV. Therefore, the previous discussion does not change in
a significant way. Secondly, $\Lambda$ is restricted by the
suggested condition $\Lambda \sim 4 \pi f$~\cite{relationlambdaf}
and, as for the scale $\Lambda$, the electroweak precision tests
seem to indicate an experimental lower bound $\Lambda \gsim 10$
TeV~\cite{barbieri}.

With those possible values of the three parameters $\lambda_T$,
$f$ and $\Lambda$, we will focus on studying some generic
information of the LH model derived from the condition
$\mu^2=\lambda v^2.$ For the numerical analysis, and by taking
into account the previous discussion, we varied the above three
parameters in the following ranges, $0.5 < \lambda_T < 2$, $10$
TeV $<\Lambda< 12$ TeV and, accordingly, $0.8$ TeV $<f<1$ TeV. Let
us first describe the behaviour of $\mu^2$ with $\lambda_T$, $f$
and $\Lambda$. In general, the $\mu^2$ corrections increase with
$f$ for $0.5 < \lambda_T < 2$, having a minimum for a certain
value of $\lambda_T$, which corresponds to a minimum for
$m_{T}$. These corrections also increase with $\Lambda$, but less
dramatically. We find that the lowest value of $\mu$ is $\mu = 0.48$ TeV,
when $\lambda_T=0.72$, $f=0.8$ TeV and $\Lambda=10$ TeV.
Notice that, even if $f$ and $\Lambda$ have the expected values
for these parameters on the LH models, the minimum possible value
for $\mu$ does not correspond to the value of this
parameter predicted by the data. We will discuss this point later on.

\begin{figure}[t]
\begin{center}\vspace{-0.3cm}
{\rotatebox{270}{\epsfig{file=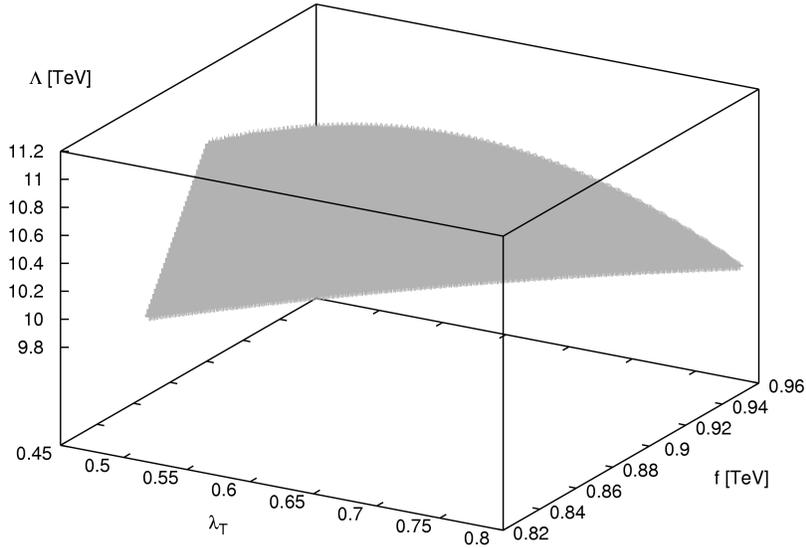,scale=0.5}}}
\caption{Values of $\Lambda$, $f$ and $\lambda_{T}$, with
$0.5 < \lambda_T < 2$, $0.8$ TeV $<f<1$ TeV and $10$ TeV $<\Lambda< 12$
TeV, which satisfy the condition
$\mu^{2}=v^{2} \lambda$, taking $0.3$ TeV $<\mu < 0.5$ TeV.}
\label{higgs}
\end{center}
\end{figure}

Once the corrections to the quartic coupling $\lambda$ has also
been computed (see~(\ref{eq:lambdaTOP})), the consistency of the
LH model is constrained by the non-trivial
condition~(\ref{eq:relation}), as already established. The result
for $\lambda$ in (\ref{eq:lambdaTOP}) has not been given so far
and, therefore, the relation between $\lambda$ and $\mu$ has not
been studied before and one would expect it to put significant
constraints on the LH models. In general, the corrections to
$\lambda$ grow with the scale $\Lambda$ but, conversely, we find
that they decrease with the symmetry breaking scale $f.$
Figure~\ref{higgs} shows the surface of solutions for this
non-trivial condition, by considering the above intervals for
$\lambda_{T}$, $f$ and $\Lambda$, and by assuming that $0.3$ TeV
$< \mu < 0.5$ TeV. Clearly, once the condition~(\ref{eq:relation})
is assumed, the allowed region of the parameter space is
considerably reduced. For $\lambda_{T}$, $f$ and $\Lambda$ in the
above intervals, there exist just few points of the parameter
space that are in agreement with a predictive model. Besides, we
stress that the lowest value of $\mu$ that
satisfies~(\ref{eq:relation}) is $\mu=0.52$ TeV, when $f=0.85$
TeV, $\lambda_T=0.52$  and $\Lambda=10$ TeV.

Finally, it is known that the mass term for the Higgs field is
generated at the one-loop level by logarithmically  divergent
diagrams. The one-loop contribution to the Higgs mass from the top
sector is given in eq.~(\ref{eq:muparameter}) ($m_H^2=2\mu^2$).
The lowest value for $\mu$, which satisfies the condition
$\mu^{2}=\lambda v^2$, is of the order of $0.5$ TeV. However, it
is well known that $\mu$ is forced by data to be at most of order
$200$ GeV. Therefore, other contributions must be included in
order
 to obtain the predicted value of $\mu.$ Notice that the
Higgs mass parameter in the LH
 model also receives contributions (not included here)
from the vector bosons sector (1-loop correction)
and from the scalar sector (2-loop correction),
which are opposite in sign to the top-fermion
contribution. We discuss in the following how
the vector boson contributions could reduce the value
of $\mu$ to its allowed value.

Concerning the gauge bosons interactions, we consider the two
different models that have been described in Section 2: the
original LH with two $U(1)$ groups ({\it{model I}}) and the other
one with just one $U(1)$ group ({\it{model II}}). Once the quantum
corrections involving gauge interactions are included, the
logarithmically enhanced contributions of the vector bosons to the
mass term of the Higgs field in each model are given, respectively,
and at one-loop level, by
\begin{equation}
\mu^{2\,\,\,{\it{I}}}_{{\mbox{g}}}
=-\frac{3}{64\pi^{2}}\left(3g^{2}M_{W'}^{2}
\log\left(\frac{\Lambda^{2}}{M_{W'}^{2}}+1\right)+g'^{2}M_{B'}^{2}
\log\left(\frac{\Lambda^{2}}{M_{B'^{2}}}+1\right) \right)\,,
\label{eq:mu2model1}
\end{equation}
\begin{equation}
\mu^{2\,\,\,{\it{II}}}_{{\mbox{g}}}= -\frac{3}{64 \pi}\left(3
  g^{2}M_{W'}^{2}\log\left(\frac{\Lambda^{2}}{M_{W'}^{2}}+1\right)+g'^{2} \Lambda^{2}\right)\,,
\label{eq:mu2model2}
\end{equation}
where $M_{W'}$ and $M_{B'}$ are the heavy gauge boson masses,
\begin{equation}
M_{W'}=\frac{f g}{2\cos\psi \sin\psi} \hspace{0.2in}\mbox{and}
\hspace{0.2in} M_{B'}=\frac{f g'}{\sqrt{20}\cos\psi' \sin\psi'}\,;
\end{equation}
with $\psi$ and $\psi'$ being the mixing angles for the $W'$ and $B'$
states. Note that the different results for these two models comes
from the fact that there is no $B'$ in {\it{model II}}.

Let us now estimate the cancellations that could
occur between the fermion sector and
the vector boson sector by keeping $\mu^{2}$ of order 200 GeV and,
therefore, $m_{H}$ light. For the numerical analysis, we varied the
$\lambda_{T}$, $f$ and $\Lambda$ parameters in the intervals given
before.

For {\it{model I}}, in order to avoid the gauge masses being too
small or too big, we impose that $0.1 < \cos \psi <0.9$ and $0.1 <
\cos \psi'<0.4$. Once cancellations occurs,
we find that the lowest value for $\mu$ is
$0.338$ TeV for $\lambda_{T}\simeq 0.7$,  $f\simeq 0.8$ TeV,
$\Lambda \simeq 10$ TeV, $\cos \psi \simeq 0.1$, and $\cos \psi'
\simeq 0.1$. On the other hand, in the case of {\it{model II}}, by
considering the same numerical values as above, we obtain better
results. The lowest value for $\mu$ is now of order of $0.2$ TeV,
with $\lambda_{T}\simeq 0.65$, $f\simeq 0.8$ TeV, $\Lambda \simeq
11.9$ TeV, and $\cos \psi \simeq 0.1.$ Therefore, we find that the
so-called {\it{model II}} is more effective than {\it{model I}}
from the point of view of  cancellations between different sectors
of the LH model.

From the above results, we could conclude
that the condition $\mu^2=\lambda v^2$ is very important in the analysis of
 predictions from the LH model, and that the inclusion
of the contributions from the vector bosons on both $\mu$ and
$\lambda$ will be crucial to increase the allowed region of the
parameter space. To explore the complete region of the parameter
space in which the LH model is consistent with the data, we plan
to make the full analysis with the inclusion of contributions from
all the sectors of the model~\cite{future}.

\section{Conclusions}

In the SM the Higgs mass $m_H$ receives quadratic radiative
corrections coming from the gauge bosons, the Higgs self-coupling
$\lambda$ and from the top quark (the latter being negative). The
requirement of not having one-loop contributions to the  squared
Higgs mass larger than $10\% $ of the tree-level value, and the
experimental constraint $115$ GeV $< m_H < 200 $ GeV, sets the SM
ultraviolet cut-off $\Lambda_{\rm{SM}}$ to be lesser than $2$ or
$3$ TeV. However, to avoid conflict with electroweak precision
observables, a scale of the order of $10$ TeV seems to be needed.
This is the so called little hierarchy problem which the LH model
pretends to solve.

In this work we have computed and analyzed the fermion
contributions to the low energy Higgs effective potential and we
have illustrated the kind of constraints on the possible values of
the LH  parameters that can be set by requiring the complete LH
Higgs effective potential to reproduce exactly the SM potential.
The analysis we have done is relevant whenever one assumes that
the new physics decouples from the low energy physics. Our results
are compatible with the LH model solving the little hierarchy
problem, but the region of LH model parameter space compatible
with this possibility probably is not very large.

We have explored the region of the $\lambda_{T}$, $f$ and
$\Lambda$ parameter space compatible with the condition $\mu^2 =
\lambda v^2$, taking $m_T \leq 2$ TeV, $\Lambda \thicksim 4 \pi
f$, and also imposing the requirement of having $\mu$ smaller than
$0.5$ TeV. The scales $f$ and $\Lambda$ run around the typical
scales predicted by the LH models. For this purpose we have
computed the fermion contributions to the quartic coupling
$\lambda$ at the one-loop level. Since the values obtained for
$\mu$ are relatively high, the inclusion of the gauge and the
scalar sector of the model is needed to reduce $\mu$ to its
expected value. Therefore, more detailed computations, including
the full one-loop gauge boson and the relevant two-loop Goldstone
boson contribution, are needed in order to establish definitely
the validity of the LH model and its compatibility with the
present phenomenological constraint including the precise form of
the Higgs potential. Work is in progress in this direction
\cite{future}.

\section*{Acknowledgments}

This work is supported by DGICYT (Spain) under project number
BPA2005-02327.
The work of S.P.\ has been partially supported by the European Union
under contract No.~MEIF-CT-2003-500030.
L.T. would like to thank Javier Almeida Linares (UCM, Spain) and
Javier Rodriguez Laguna (SISSA, Italy) for their
valuable guidance to C$++$ programming.

\end{document}